\documentclass{llncs}

\usepackage{llncsdoc}
\usepackage{url}
\usepackage{graphicx,epsfig}% Include figure files
\usepackage{subfig}
\usepackage{graphics,dcolumn,bm,epic,eepic,float}
\usepackage{amssymb,amsmath,multirow,rotate,color}
\usepackage{epstopdf}
\usepackage{soul}
\usepackage[draft]{hyperref}
\usepackage{color}
\usepackage[]{algorithm2e}
\usepackage{booktabs}
\usepackage[russian,ukrainian,english]{babel}
\usepackage[T1,T2A]{fontenc}
\usepackage[utf8]{inputenc}

\newcommand\reffig[1]{Fig.~\ref{#1}}

\newcommand\reftab[1]{Table~\ref{#1}}

\newcommand{\specialcell}[2][c]{%
 \begin{tabular}[#1]{@{}c@{}}#2\end{tabular}}

\usepackage{footnote}
\makesavenoteenv{tabular}
\makesavenoteenv{table}

\begin{document}

\title{Identifying Partisan Slant in News Articles and Twitter during Political Crises}

\author{Dmytro Karamshuk$^1$, Tetyana Lokot$^2$, Oleksandr Pryymak$^3$, Nishanth Sastry$^1$}
\institute{$^1$ King's College London $^2$ Dublin City University $^3$ Facebook \\
\email{$^1$\{firstname.lastname\}@kcl.ac.uk, $^2$tlokot@umd.edu , $^3$ opr@fb.com}
}

\maketitle 

\begin{abstract}

In this paper, we are interested in understanding the interrelationships between mainstream and social media in forming public opinion during mass crises, specifically in regards to how events are framed in the mainstream news and on social networks and to how the language used in those frames may allow to infer political slant and partisanship. We study the lingual choices for political agenda setting in mainstream and social media by analyzing a dataset of more than 40M tweets and more than 4M news articles from the mass protests in Ukraine during 2013-2014 — known as "Euromaidan" — and the post-Euromaidan conflict between Russian, pro-Russian and Ukrainian forces in eastern Ukraine and Crimea. We design a natural language processing algorithm to analyze at scale the linguistic markers which point to a particular political leaning in online media and show that political slant in news articles and Twitter posts can be inferred with a high level of accuracy. These findings allow us to better understand the dynamics of partisan opinion formation during mass crises and the interplay between mainstream and social media in such circumstances.

\end{abstract}

\section{Introduction}

Social media have become a crucial communication channel during mass political or civic events by shaping "a civic and democratic discourse in a vacuum of opportunities" \cite{howard2010digital}. As academic debates contest the nature of social media as an alternative public sphere \cite{papacharissi2009virtual}, it is important to study the \emph{interrelationships between mainstream media and social networks} in shaping public opinion during mass protests, especially in regards to the origin and dissemination of news frames \cite{oates2013twilight}. It is also of interest to consider how \emph{propaganda and manipulation} in the information sphere work: where partisan language and frames originate, how they spread, and whether there are certain markers that would allow to trace the distribution and paths of such content.

In this study, we analyze the role of social and mainstream media in shaping and disseminating partisan content frames during social unrest and crisis. Specifically, we focus on the mass protests in Ukraine during 2013-2014 - known as \emph{"Euromaidan"} - in which social media played a remarkable role, helping to raise awareness, cover, and discuss ongoing events; and the post-Euromaidan Russian occupation of Crimea and the conflict between Russian, pro-Russian and Ukrainian forces in eastern Ukraine (2014-2015), periods that were characterized by a parallel information and propaganda war occurring in mainstream media and online together with military action on the ground.

We explore the extent to which lingual choices in online discourse can illuminate the \emph{partisan confrontation} between political factions during mass crises through the analysis of the two complementary datasets of Twitter posts and news articles. More specifically, we exploit natural language processing to single out language that points to a particular political leaning and to observe whether these markers are detectable in both mainstream media language and social media posts at scale. Our contributions can be summarized as follows:

\begin{itemize}
	\item We exploit the word embedding approach~\cite{mikolov2013distributed,mikolov2013efficient} to identify the indicators of partisan slant in news articles and validate it over the text corpora of around 4M news articles collected during the Ukrainian conflict. Our analysis reveals a \emph{strong use of highly polarized partisan content frames} in news articles on both sides of the conflict. 
	\item Next, we design a machine learning approach for \emph{detecting the markers of partisan rhetoric} in news articles with minimal efforts required for supervision. This is achieved by a "coarse-grained" labeling of the articles based on the partisan slant of the news agencies they originate from. Our approach - trained on a collection of articles from 15 representative news agencies -	is able to achieve $60$--77\% accuracy in distinguishing between the news articles with pro-Ukrainian, Russian pro-government and Russian independent slants during the Ukrainian conflict.
  \item Finally, we study the inter-relation between traditional and social media during conflicts through an analysis of individual news sharing patterns among Twitter users and find that most of the users are \emph{exposed to a variety of news sources but with a strong partisan focus}. Using our machine learning approach, we are also able to predict the partisan leaning of Twitter users from the content of the tweets with an accuracy up of 70\%. 
\end{itemize}

In summary, we demonstrate that studying the lingual choices of Twitter users and news media adds a new dimension to understanding the dynamics of information flows and partisan idea dissemination in the space between social networks and mainstream media. We also demonstrate that lingual choices-based machine learning models can be highly effective at automatically predicting the political slant of mainstream media and Twitter users, which can have serious implications for political expression in repressive and authoritarian regimes. 

\vspace{-1.1em}
\section{Related Literature}
\vspace{-1.1em}

Computational social scientists have given substantial attention to the mainstream and social media activity around political and social change, and to the role information shared on these platforms plays in influencing political and social agendas, protest movements and events, and public opinion. Researchers have explored the role of social media and mainstream media actors in information diffusion and protest message amplification in networks through the prism of the \emph{collective action theory} \cite{gonzalez2013broadcasters}, as well as the role of social networks in recruitment and mobilisation during protest, revealing connections between online networks, social contagion, and collective dynamics \cite{gonzalez2011dynamics}. 

A broad swathe of quantitative studies have focused on determining the factors that influence political leanings of social network users and metrics that allow to classify and predict this kind of political bias. Several studies have considered the \emph{predictive power} of political news sharing habits on Twitter \cite{an2014sharing}, the influence of partisan information sharing on political bias among Facebook users \cite{an2014partisan} and Twitter actors \cite{hegelich2016social,conover2011predicting}, and compared differences and biases in news story coverage, dissemination, and consumption among online mainstream and social media \cite{chakraborty2016dissemination}. Others have noted the difficulty of connecting selective exposure to political news on Facebook to partisanship levels of users \cite{an2013fragmented}. At the same time, researchers suggest that analysis of information consumption and distribution habits of social network users does provide data on media exposure, the relationship between various classes of media, and the diversity of media content shared on social networks \cite{an2011media}. 

Some studies have noted that reliably inferring the political orientation of Twitter users and generalizing the findings is notoriusly difficult due to it being one of the ``hidden'' attributes in social network data and due to differences between politically active groups of users and the general population \cite{cohen2013classifying}. However, other academics suggest that studying some relationships on social networks, like the co-subscriptions relationships inferred by Twitter links \cite{an2012visualizing}, allows for some understanding of the underlying media bias — and subsequently, political bias — of social network users. Another study showed that applying machine learning techniques to classify political leanings on Twitter based on political party messages can reveal partisanship among users \cite{boutet2013s}. A number of studies present a comparison between the predictive power of the users' social connections and their content sharing patterns for inferring political affiliation, ethnicity identification and detecting affinity for a particular business \cite{pennacchiotti2011democrats,pennacchiotti2011machine}.

While the research described above uses a fairly large spectrum of methods to study, classify and find connections between social media users' behavior and their media and political preferences, most of the studies referenced employ social network analysis or related methods, focusing on relationships between actors or their behaviors within the network, such as sharing links, following other actors, etc. More recent studies have used computational methods to assess forms of political organization on social media \cite{aragon2016movement}, employed machine learning models to classify rumor and misinformation in crisis-related social media content \cite{zeng2016unconfirmed}, and used deep neural networks to identify and analyze election-related political conversations on Twitter on a longitudinal scale \cite{vijayaraghavan2016automatic}.

We propose augmenting these approaches with a focus on the \emph{linguistic variables} present in the data, and using natural language processing and machine learning techniques to gain further insight into how political and ideological messages travel between mainstream and social media, and how these lingual choices reflect the partisan nature of mainstream media outlets and, subsequently, social media users and their content consumption and sharing habits. Such an approach would allow for a more granular understanding of how language changes allow to detect both important events and partisan leanings in mainstream and social media data.

\vspace{-1em}
\section{Background and Datasets}
\vspace{-1em}

The Euromaidan protests and  subsequent political crisis are the outcome of a continuing trend in the post-Soviet arena. The last decade or so has seen an increase in mass protest actions in the region, with protests erupting in Russia, Belarus, Ukraine, etc. From the 2004 \emph{Orange Revolution} in Ukraine to the \emph{Bolotnaya rallies} of 2011-2012 in Moscow to the Euromaidan protests in Ukraine, a gradual increase in the use of digital technologies and media platforms by citizens has become evident \cite{goldstein2007role,oates2013twilight,onuch2015euromaidan}. At the same time, the region is characterized by a problematic media climate, with mainstream media often co-opted or controlled by the state or the oligarchy. The interplay and mutual influence of mainstream and social media emerge as crucial for understanding the political and civic developments in the region and thus demand more scholarly attention. 

\begin{figure*}
	\vspace{-1em}
	\hspace{-1em}\includegraphics[width=1.1 \textwidth]{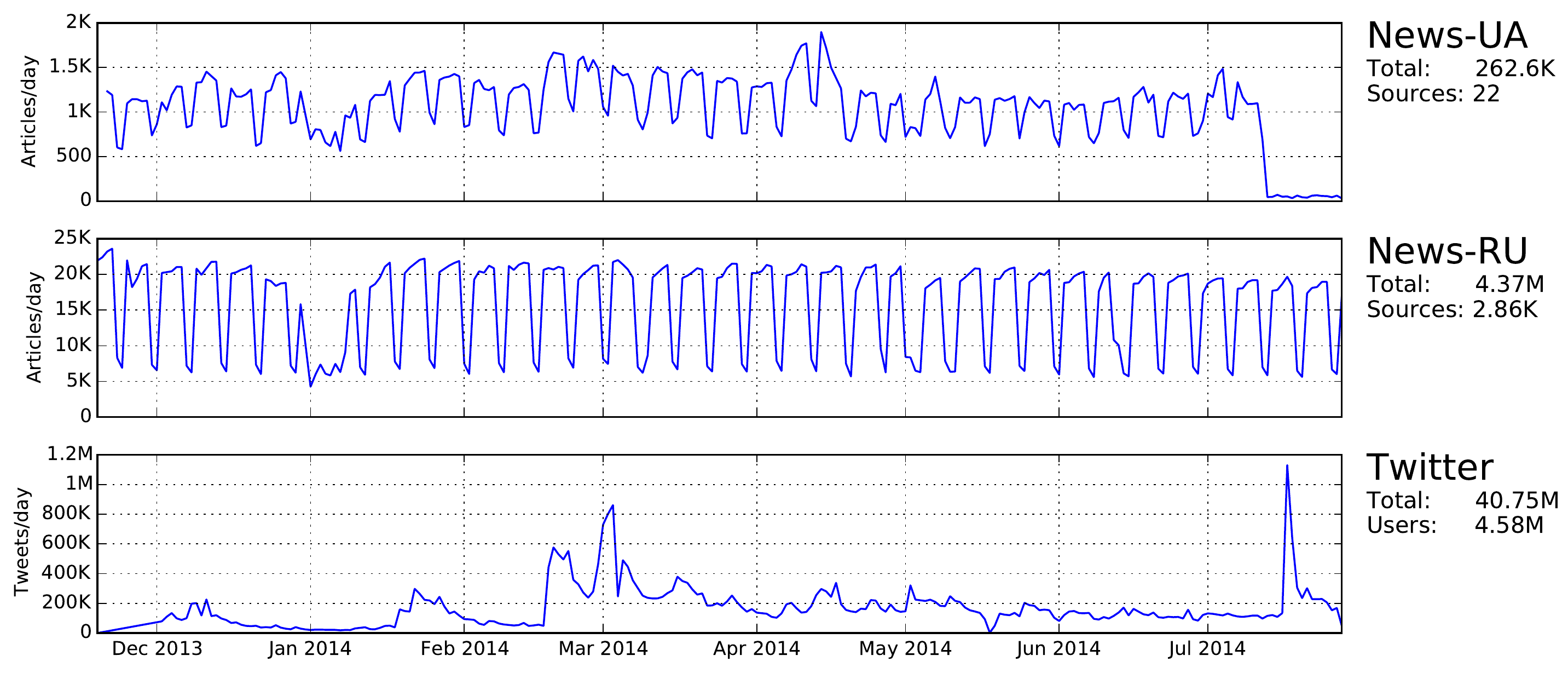}
	\vspace{-1em}
	\caption{\textbf{Description of the Datasets.} The Twitter dataset was collected via \emph{Twitter Streaming API} (spikes in daily volume correspond to higher discussion volumes around Ukraine-related topics, February-March is the most active phase of the protest, spike in July corresponds to the MH17 airplane tragedy); the News-RU dataset was crawled from the \emph{news.Rambler.ru} news aggregator website (periodic pattern reflects the weekly cycle); and the News-UA dataset was provided by the developers of the \emph{Kobzi} mobile app.}
\vspace{-1em}
	\label{tab:dataset-description}
\end{figure*}

A number of researchers have already examined some of the more general aspects of Euromaidan, such as the reasons for the protest \cite{diuk2014euromaidan}, who the protesters were \cite{zelinska2015were}, how the protest came together and evolved \cite{onuch2015euromaidan}. However, few have investigated the use of civic media by Euromaidan participants beyond simply saying that social media were used in the protest as `tools' for mobilization and information dissemination \cite{onuch2015euromaidan}. A deeper and more large-scale analysis of the relationship between mainstream media coverage of the crisis and the grassroots social media data around the political unrest, enabled by computational and big data tools and complemented by qualitative analysis, could reveal more about how the partisan agenda during the protests was formed and transformed through lingual choices and memorable memes, and who was able to exert influence on the lingual frames used by the multitudes of social media users and media outlets. Such investigation could also shed light on the reasons and mechanisms of the information war between Ukraine, Russia, and the West that gained in scale after the Euromaidan protests shifted into the crisis characterized by Russia’s annexation of Crimea and the pro-Russian uprising in eastern Ukraine.

To research these questions in this paper we analyze two complementary datasets: A \emph{social media} dataset which consists of over $40$M tweets collected for the three most prominent hashtags during and after the Euromaidan protests - \#euromaidan \#ukraine \#kyiv (and their Ukrainian and Russian equivalents) via Twitter Streaming API and a dataset of more than $4$M \emph{news articles} collected from a large Russian news aggregator (Rambler.ru) and its Ukrainian counterpart (smartphone app Kobzi). All three datasets were collected in the period between December 2013 and July 2014. The parameters of the datasets are summarized in \reffig{tab:dataset-description}.

\section{Exemplar indicators of partisan slant}

\subsection{Methodology}

Our initial interest in exploring the \emph{lingual choices} made by mainstream media sources and social media users was sparked by observing the emergence of a number of memes and buzzwords introduced during the events. In this section, we aim to measure the presence of these keywords in the rhetoric of the parties involved in the conflict and explore the extent to which this analysis can be automated. 

To analyze the lingual choices of online media sources during the Ukrainian conflict we first exploit the \emph{word embedding} methodology proposed in \cite{mikolov2013distributed,mikolov2013efficient}. The proposed approach devises vector representations of words by analyzing the textual context in which they appear. This is achieved by training a model which for a given word, represented by a vector $X_i$, aims to infer the most likely $2 \times j$-surrounding words vectors which constitute the lingual context where the word was used~\footnote{We used the Skip-Gram model as it provided more interpretable results.}, e.g, $f(X_i) = (X_{i-j}, ..., X_{i-1}, X_{i+1}, ..., X_{i+j})$. Intuitively, semantically closed words are expected to appear in similar contexts and so should produce similar outcomes when applied as the arguments of the function $f(X)$. Thus, if trained on a significantly large text corpus the algorithm is able to assign close-by vectors to the words with similar meanings, thereby providing a powerful framework for analyzing the semantics of the word choices.

Equipped with the Word2Vec implementation\footnote{Gensim natural language processing library https://radimrehurek.com/gensim/}, we build word representations for the textual corpora extracted from our datasets. Note that we mainly focus on vector spaces extracted from the news corpora (News-UA and News-RU) as Twitter's limit of $140$ characters significantly constrains the applicability of this approach. In pre-processing, we remove rare words with less than $10$ occurrences in each of the corpora and end up with a dictionary of $87$K words trained from a content corpus of $600$M words. Note that while the Ukrainian media space is generally bilingual, in our analysis we only focus on Russian-language news articles, in order to allow for a fair cross-comparison of the results, i.e., we focus on the Russian-language media sphere as it presents a sufficient diversity of news sources, with political views spread across the spectrum, ranging from pro- to anti-Kremlin and from pro- to anti-Ukraine, including both Ukrainian and Russian media outlets.

\subsection{Mining semantics of word choices}

In \reftab{tab:word2vec-examples} we present several examples of word associations which were mined from both our news corpora. We pick several loaded terms which were the prominent indicators of partisan rhetoric during the conflict and match these words to the most similar ones according to the trained Word2Vec dictionaries (as measured by the cosine similarity). The results illustrate that the trained model corresponds to our understanding of the semantics of chosen terms. 
For instance, the word \emph{referendum (референдум)}, frequently used in the context of the Russian-backed annexation referendum in Crimea, lies very close to its synonyms, e.g., `plebiscite', `voting', etc., in the devised vector space in both the News-RU and News-UA corpora, from both sides of the conflict.

On the other hand, we also notice that some of the synonyms discovered by Word2Vec reflect the propagandistic rhetoric of official Russia and Ukraine during the conflict. For instance, in the News-UA corpus (but not in the News-RU corpus), the word \emph{referendum (референдум)} is associated with the words `non-legitimate' and `fake referendum', and reflect the official Ukrainian state's position on the plebiscite. Similarly, the word \emph{aggressor (агрессор)} as used in association with Russia's actions in eastern Ukraine and Crimea is associated with the words `cynical' and `unprovoked' and captures the attitude of the Ukrainian government in regards to the events.
 
A similar partisan rhetoric is also observed in Word2Vec word representations mined from the Russian news corpus (News-RU). For instance, we observe that the loaded term \emph{junta (хунта)}, which was extensively used by some Russian media sources to demonize the transitional government formed in Ukraine after the Euromaidan protests, is strongly associated with the words `neo-fascist' and `pro-Ukrainian' in the Russian news corpus (News-RU). Similarly, the word \emph{punisher (каратель)} which was used by some Russian media sources to label the Ukrainian Armed Forces and their operations in the conflict in eastern Ukraine is closely associated with `fascism' and `terrorize', as well as with Stepan Bandera, a controversial figure who has been acknowledged by some in the new Ukrainian government.

Such biased lingual choices are in alignment with recent findings that politically charged rhetoric and biased language were central to the discourse around the Euromaidan protests in Ukraine and the subsequent conflict in eastern Ukraine, both of which featured interference by Russian political forces~\cite{szostek2014media,zhukov2016reporting}.

\vspace{-1em}
\begin{table}[!htbp]
\centering
\begin{tabular}{ | l | c | l | c |}
\multicolumn{4}{c}{\large \textbf{News-RU}} \\ \hline
\multicolumn{2}{|c|}{\textbf{каратель [punisher]}} & \multicolumn{2}{c|}{\textbf{хунта [junta]}} \\ \hline
бандера [Bandera~\footnote{Stepan Bandera - the leader of the western Ukrainian resistance during World War II, accused by Soviets of collaborating with Nazis.}]  & 0.79 & восстание [uprising] & 0.76 \\
безоружный	[unarmed] 	& 0.79 & проукраинский [pro-Ukrainian] & 0.72 \\
расстреливать [shoot]	& 0.77 & неофашист [neo-fascist] & 0.72 \\
терроризировать [terrorize] & 0.77 & неонацист [neo-Nazi] & 0.71 \\
оккупант [invader] & 0.76 & путчист [coupist] & 0.71 \\
фашист [fascist] & 0.76 & самозванец [imposter] & 0.71 \\ \hline
\multicolumn{4}{c}{} \\

\multicolumn{4}{c}{\large \textbf{News-UA}} \\ \hline
\multicolumn{2}{|c|}{\textbf{референдум [referendum]}} &\multicolumn{2}{c|}{\textbf{агрессор [aggressor]}} \\ \hline
псевдо-референдум [fake referendum] & 0.72 & неспровоцированный [unprovoked] & 0.72\\
плебисцит [plebiscite] & 0.71 & враг [enemy] & 0.70\\
голосование [voting] 	&  0.67 & циничный [cynical] & 0.66 \\
автономный [autonomous] & 	 0.66 & развязанный [launched] & 0.66\\
нелегетимный [non-legitimate]	& 0.64& извне [external] & 0.65\\
присоединение [attachment]& 0.63 & оккупационный [invasive] & 0.65\\
отсоединение [separation] &  0.62 &	террор [terror] & 0.64\\ \hline
\end{tabular}
\vspace{2em}
\caption{\textbf{Examples of word embeddings mined from the news datasets during the Ukrainian conflict.} The table presents the word associations mined using the Word2Vec algorithm for several loaded terms (bold font) in Russian (top) and Ukrainian (bottom) news.} 
\label{tab:word2vec-examples}
\vspace{-3em}
\end{table}

\vspace{-1em}
\section{Identifying Slant of News Stories}

Inspired by the observations from the previous section we next question the power of the word choices to characterize the difference in partisan media at scale, i.e. are the linguistic choices of partisan media substantially different such that we can automatically differentiate them? We address this question by developing a classification machine learning model and conducting an extensive validation of the model over our news datasets. 

\subsection{Methodology}

Using the Word2Vec word representations from the previous section, we train a supervised learning algorithm to find the best indicative words which characterize the language style of a given party. To this end, we first manually classify the Top-30 most popular Russian news agencies\footnote{as ranked by the Medialogia rating agency http://goo.gl/JNvx0Y} as having a strong pro-government or opposition slant and complement this list by the Top-5 Russian-language sources from the Ukrainian internet segment. We achieve this by manually examining 20 or more articles per each news source for qualitative signs of slant, and by investigating public ownership records and publicly available info about the media outlets, their owners and affiliations. While granular and manual, such an approach can be replicated in other studies using media sources, as public ownership records are usually available and provide enough context for classification, while manual examination for signs of slant is based on a designated set of relevant keywords. We also remove all neutral sources from our analysis, i.e those that have not indicated a particular partisan slant. Finally, we only consider news articles related to the Ukrainian unrest and conflict\footnote{This is achieved through filtering the corpora by the relevant keywords, i.e., "kyiv", "ukraine", "donbass", "maidan", "crimea", "luhansk", "dnr" and "lnr". Adding a wider set of keywords had little effect on improving the recall of filtering.}. Our classification results in three categories of media outlets: UA, RU-ind, and RU-gov, exhibiting pro-Ukrainian, Russian-independent and Russian-pro-government points of view on the Ukraine unrest and conflict respectively.

Next, we cluster the Word2Vec vectors obtained from the previous section, and use the produced clusters as a feature space to describe the content of each article. In more detail, we apply k-means clustering with $N = 1000$ clusters to the word vectors of the combined News-UA and News-RU corpus. Then, for each article we calculate a $1000$-items long feature vector $X$ that represents the normalized frequencies of occurrences of words from each cluster that occur in the article, and train a function $f(X) \rightarrow \{RU-gov, Ru-ind, UA\}$ to identify the partisan slant of the article (e.g., whether coming from Russian pro-government (Ru-gov), Russian independent (RU-ind) or Ukrainian (UA) news source\footnote{We refer to each of these three classes as `party' or 'parties' in the rest of this paper.}). 

\vspace{-1em}
\begin{table}
\centering
\begin{tabular}{ | c | c | c | c |}
\hline
\textbf{word} & \textbf{$\rho_w^s$} & \textbf{$\bar{\rho_w^s}$} & \textbf{News source} \\ \hline
ъ [Ъ] & 0.38 & 1.0 & Ъ-Kommersant \\
m24.ru [m24.ru] & 0.30 & 1.00 & Moscow 24\\ 
известиям [izvestiam] & 0.31 & 0.93 & Izvestia \\
господин [gentlemen] & 0.29 & 0.79  & Ъ-Kommersant \\ 
подробнее [more] & 0.36 & 0.69 & Ъ-Kommersant \\
рбк [RBC] & 0.27  & 0.68 & RBC\\ 
\hline
\end{tabular}
\vspace{0.5em}
\caption{\textbf{Exemplar markers of individual news sources} as identified by the words with the highest relative frequencies $\bar{\rho_w^s}$ across all words and news sources.}
\label{tab:source-bias}
\vspace{-3.5em}
\end{table}

\subsection{Reducing news source bias}

One crucial aspect to account for in the proposed approach is its ability to learn linguistic patterns that generalise across all news sources of a given partisan slant, rather than markers of individual news outlets. Since the training data in the above method is sourced from a selection of few news sources, without a generalisable approach, the classifier could simply learn to label the partisan slant of an article based on unique words, or \emph{news source markers}, that are specific to a particular biased news source.  

For example, it is common that the name of the news source or its correspondents are explicitly mentioned in the byline or text of the article, making it easily distinguishable among other texts. Supposing the training data contains a biased Russian pro-government news source $B$ whose name appears in every article from $B$, we might learn a model that the word $B$ is indicative of a Pro Russian-government partisan slant. Although this is useful to classify other articles from $B$, it does not help identify other pro-russian news sources or articles. We therefore need to adapt the method to learn labels that generalise to all news sources by ignoring news source markers.

\vspace{-1em}
\subsubsection{Description of the problem.}

To show that news source markers are indeed widespread among the news articles in our dataset, we measure the relative frequencies of words appearing in articles from each individual news source. More formally, we define the frequency $\rho_w^s = \frac{N_w^s}{N_s}$ of word $w$ in news source $s$ as a share of all articles $N^s$ from news source $s$ in which word $w$ appears at least once and compare it to the sum\footnote{Note that this is equivalent to using the mean $\frac{\sum_{s \in S}{\rho_w^s}}{|S|}$, since the sum for all words is computed over the same set of sources $S$.} $\sum_{s \in S}{\rho_w^s}$ of the observed frequencies of $w$  across all news sources $s \in S$, i.e., we define the ratio $\bar{\rho_w^s} = \frac{\rho_w^s}{\sum_{s \in S}{\rho_w^s}}$ to identify words which are highly unique to particular news sources.

The top news source markers are extracted as the words with the highest ratio $\bar{\rho_w^s}$ across all words and all news sources.  Table~\ref{tab:source-bias} shows the top few markers. As expected, we observe that articles from some of the news sources - such as Ъ-Kommersant, Moscow 24, Izvestia, and RBC - contain very vivid word markers of that news source. For instance, the words "Ъ" and "m24.ru" used as abbreviations of the Ъ-Kommersant and Moscow 24 news papers appear only within the news articles originating from these two sources. More interestingly, we observe a very high relative frequency of mentioning other general  words such as "господин [gentlemen]" and "подробнее [more]" in articles originating from Ъ-Kommersant, suggesting that there might be other word markers -- beyond just the obvious names of newspapers -- which indicate the writing style of an individual news source.

\vspace{-1em}
\begin{table}
\normalsize
\centering
\begin{tabular}{ |l | c | c | c |}
 \hline
  \textbf{Metric} & \specialcell{\textbf{RU-gov} vs. \\ \textbf{RU-ind}} & \specialcell{\textbf{RU-gov} vs. \\ \textbf{UA}} 
			& \specialcell{\textbf{RU-gov} vs. \\ \textbf{RU-ind} vs. \textbf{UA}} \\ \hline
  Precision & 0.66 & 0.78  & 0.57 \\ \hline
  Recall & 0.65 & 0.77 & 0.60 \\\hline
  Accuracy & 0.65 & 0.77 & 0.60 \\\hline
\end{tabular}
\vspace{2em}
\caption{\textbf{Predicting partisan slant in news articles.} The results of a supervised machine learning experiment to identify whether news articles were published by either a Ukrainian (UA), a Russian pro-government (RU-gov) or Russian independent (RU-ind) news agency. }
\label{tab:articles-prediction-performance}
\vspace{-3em}
\end{table}

\vspace{-1em}
\subsubsection{Suggested solution.}
To eliminate the aforementioned news source bias in our prediction model we develop the following approach\footnote{We note that a straightforward approach of removing the most prominent news source markers -- as measured by the relative frequency introduced in the previous section -- has proved to be inefficient for the considered classification problem. In contrast, the method we introduce in the rest of this section provides a more nuanced approach in estimating the relevance of each classification feature.}. We use Random Forest classifiers known for a good performance on modeling high dimensional data and modify the underlying mechanism for constructing individual decision trees. By default, the trees of the Random Forest algorithm are constructed via a greedy search for the optimal split of the training data $D$ which minimizes the entropy of the label classes (i.e., parties in the conflict), i.e., 

\vspace{-1em}
\begin{align}
\underset{split}{\text{minimize }} H_t(D_{left}) + H_t(D_{right})
\end{align} 

where the entropy $H_t(D) = - \sum_{t \in T}{\rho_t^D log(\rho^D_t)}$ is defined on the label spaces $t \in T$ in the left $D_{left}$ and the right $D_{right}$ branches of the split ($\rho^D_t$ indicates the share of instances of class $t$ in the dataset $D$). In principle, an optimal split by this definition may be found around the word markers specific to individual news outlets (e.g., the one from Table~\ref{tab:source-bias}). To penalize this unwanted behavior of the algorithm  we introduce the entropy of a news source $s$ as $H_s(D) = - \sum_{s \in S}{\rho_s^D log(\rho^D_s)}$ which - unlike the entropy defined on labels $H_t(D)$ - characterizes the purity of the split in terms of news sources $s$ rather than parties $t$ (i.e., $\rho^D_s$ indicates the share of instances from news source $s$ in the dataset $D$). Intuitively, we aim for a split that discriminates by the party $t$ but not by the news source $s$ and, so, we aim to find the split that minimizes the entropy $H_p(D)$ while preserving a high entropy of individual news sources within the party $H_s(D)$, i.e.:

\vspace{-1em}
\begin{align}
\underset{split}{\text{minimize }} H_t(D_{left}) + H_t(D_{right}) - \alpha (H_s(D_{left}) - H_s(D_{right}))
\label{eq:entropy}
\end{align} 

where $\alpha$ is a constant controlling the effect of the proposed adjustment\footnote{The proposed approach was implemented by adapting the internal implementation of the Random Forest algorithm from the open source scikit-learn libary.}. 

\vspace{-1em}
\subsection{Cross-source validation}

In summary, we represent each article as a 1000-long feature vector $X$ based on relative word frequencies corresponding to N=1000 clusters induced by a Word2Vec representation of the entire news corpus. We then learn a function $f(X) \rightarrow \{RU-gov, Ru-ind, UA\}$ that labels the partisan slant (Russian pro-government (Ru-gov), Russian independent (RU-ind) or Ukrainian (UA) news source), whilst at the same time ensuring  (using Eq.~\ref{eq:entropy}) that the model  $f(\cdot)$ generalises beyond learning to distinguish markers specific to individual  sources. 

To properly validate the proposed approach we conduct a \emph{cross-source validation} as follows. From the news agencies labeled in the previous step we select all with at least 329 articles resulting in a dataset of five agencies from each party. The number of articles from each news agency is balanced by down-sampling the over-represented agencies. We further conduct a five-fold cross validation such that at each step we pick four news agencies from each class to train the classifier and use the remaining one for testing. Since we test the model over a news agency which has not been used for training we are able to control for over-fitting to the writing style and markers of specific outlets. 

In \reftab{tab:articles-prediction-performance} we report the average values of accuracy, precision, and recall of the proposed cross-source validation. The results in the table indicate a strong prediction performance of the algorithm (accuracy of $77$\%) in classifying the content of the news articles as coming from Russian pro-government or Ukrainian news sources. This result confirms a sharp difference in the linguistic choices characteristic for the content of Russian and Ukrainian news articles as observed in Table~\ref{tab:word2vec-examples} of the previous section. More interestingly, the accuracy is also high (i.e., 66\%) for the more difficult problem of distinguishing between Russian pro-government and Russian independent news sources which often shared a common view on individual episodes of the conflict (e.g., the annexation of Crimea). For the more general problem of discriminating between all three groups of news agencies, the algorithm is able to achieve an accuracy of 60\%. Note that this is significantly better than a na\"{\i}ve baseline of randomly guessing between the three classes (with expected accuracy of 33\%) signifying the presence of a sharp partisan slant in traditional media.

\vspace{-1em}
\section{Understanding partisan slant in Twitter}

Having studied the difference in linguistic choices characteristic for news agencies during the Ukrainian conflict we now switch to the analysis of the related discourse in social media. The focus of our analysis is on understanding the interrelation between the level of exposure to different news sources among Twitter users and the linguistic choices in their posts. 

\subsection{Methodology}

To analyze the level of exposure to various news sources among Twitter users, we rely on an established approach in the literature~\cite{an2014sharing,boutet2013s} and look at the news sharing patterns. To this end, we identify $248$K Russian-language tweets which retweet or mention articles from one of the $Y = 22$ most popular news agencies classified in one of the three groups in our dataset (e.g., whether coming from Russian pro-government (RU-gov), Russian independent (RU-ind) or Ukrainian (UA) news sources). Next, we measure the user focus on a partisan media and a specific news agency by computing the share of the articles he/she shared from his/her most preferred agency/party, correspondingly. We note that a similar approach has been previously used to analyze geographic bias of content access in social media~\cite{brodersen2012youtube}.

More formally, for each user, we measure the number of times $n_y$ she/he shared an article from a news source $y \in Y$ and calculate the user's \emph{news focus} as the share of all times he/she has shared any news article from any news source in $Y$, i.e.,
$\beta = max_{y \in Y}\left(\frac{n_{y}}{\sum_{y \in Y}{n_{y}}}\right)$.
The larger $\beta$ is, the larger the fraction of the user's shares that come from a single source. 
%\begin{align}
%\beta = max_{y \in Y}\left(\frac{n_{y}}{\sum_{y \in Y}{n_{y}}}\right)
%\end{align}
Similarly, we measure the user's \emph{party focus} $\beta_{party}$ as the fraction of news articles shared from one of the three ``parties'':  Russian pro-government (RU-gov), Russian independent (RU-ind) or Ukrainian (UA) news sources. Finally, we measure the \emph{diversity of news sources} with which the user expressed alignment by computing the cardinality of the subset $Y_{n_y > 0}$ of all sources in $Y$ from which a user has shared at least one article, i.e., $\gamma = |Y_{n_y > 0}|$.

%\begin{align}
%\gamma = |Y_{n_y > 0}|
%\end{align}\vspace{0.5mm}

\vspace{-1em}
\subsection{Patterns of news sharing}

\begin{figure*}
 \centering
  \includegraphics[width=0.32 \textwidth]{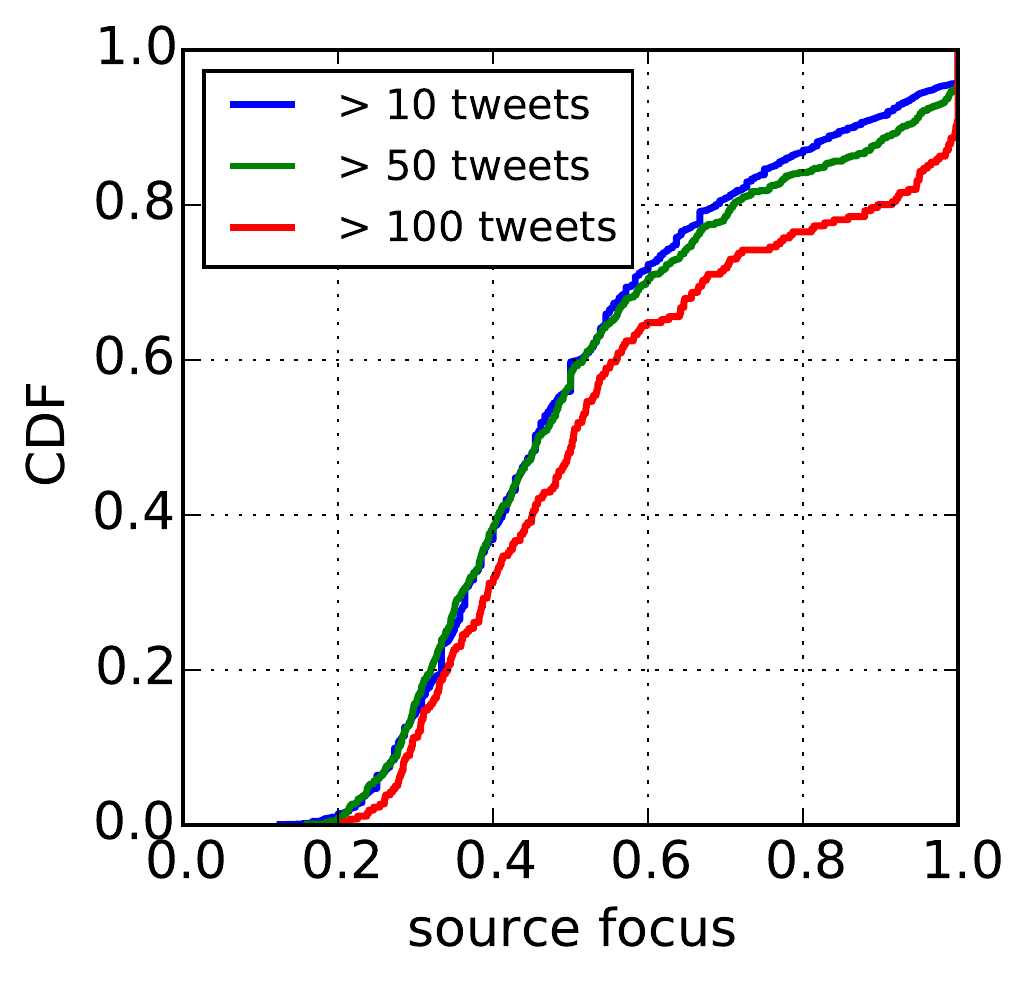}
  \includegraphics[width=0.32 \textwidth]{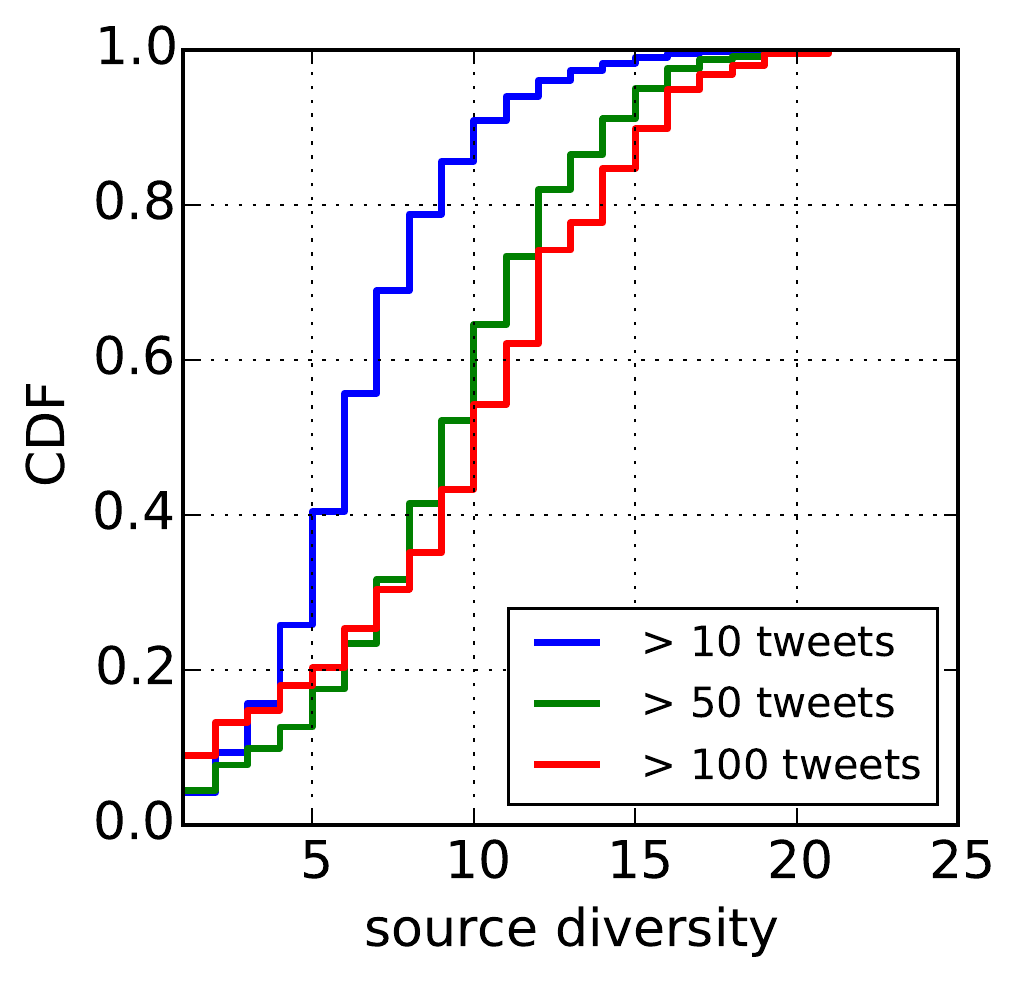}
  \includegraphics[width=0.32 \textwidth]{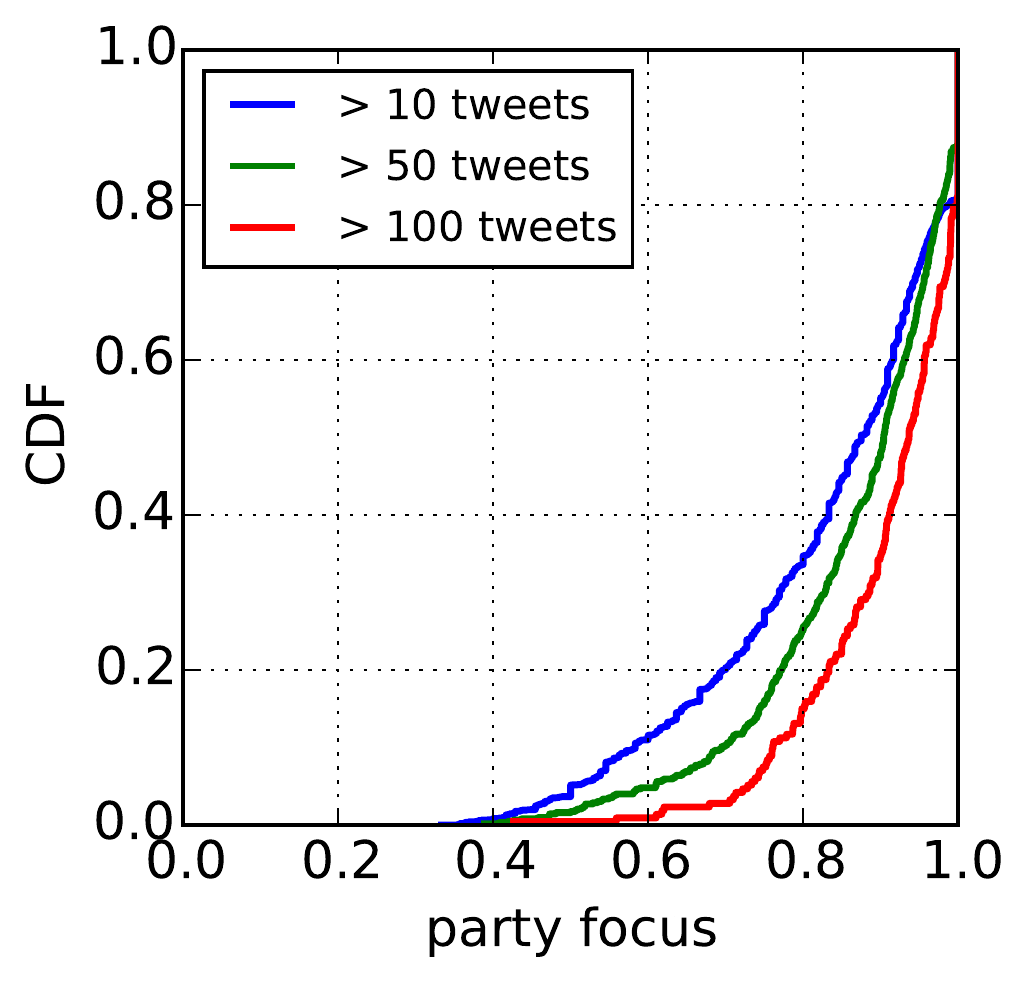}
  \caption{\textbf{Partisan news focus among Twitter users.} Distributions of (left) the user focus on preferred news sources; (middle) the diversity of news sources a user has expressed alignment with; (right) the user focus on partisan media.}
  % (e.g., whether coming from Russian pro-government (RU-gov), Russian independent (RU-ind) or Ukrainian (UA) news source).}
  \label{ref:fig-twitter-partisanship}
\vspace{-3em}
\end{figure*}

The results of the analysis are presented in \reffig{ref:fig-twitter-partisanship}. Firstly, we note that for the majority of users, less than half of all their shares come from a single news source (\reffig{ref:fig-twitter-partisanship}, left) and that the majority of users express alignment with more than $6$ news sources (\reffig{ref:fig-twitter-partisanship}, middle). At the same time, we note that although more active users tend to focus more of their shares on a single news source, they also (occasionally) share a bigger number of news sources (as indicated by the higher diversity values for the users with more than 50 and 100 tweets in \reffig{ref:fig-twitter-partisanship}.middle). 

However, this diversity of news sources is not seen when we look at the more coarse-grained picture at the level of 'parties'  (\reffig{ref:fig-twitter-partisanship}, right):  users' shares tend to be heavily focused on just one party -- on average, more than $85$\% ($90$\% for the very active users) of the shares of a user are for news sources in alignment with that user's main `focus' party. In other words, although users exhibit sophisticated behaviours such as a relatively high level of diversity in sharing from multiple news sources, most of these sources have a single partisan alignment, whether Independent Russian, Pro-Russian, or Pro-Ukrainian Government. Furthermore, both the variety of news sources and the partisan focus increase as user activity levels (number of tweets made) increase.

\vspace{-2em}

\subsection{Predicting political slant from Twitter posts}

Next, we attempt to draw the link between news exposure among Twitter users - as inferred by the news articles they share - and the language they use in their posts. From the results of the analysis in the previous section, we build on the fact that the vast majority of users have a strong partisan focus in the article they share and use that as a label of their political slant. To understand whether and to which extent the language choices in Twitter posts can indicate the political slant of the users we formulate a machine learning classifier to infer the latter from the former and validate it over our Twitter dataset. 

To this end, we focus on the $5.9$K users with at least $5$ tweets, at least $3$ of which contain news sources and balance the dataset by choosing an equal number of users from each party. To predict political slant of a Twitter user we formulate a supervised machine learning problem where we model language in a user's tweets with a feature vector $X$, using a methodology similar to the one described in the previous section, and train binary classifiers $f(X) \rightarrow \{RU-gov, RU-ind\}$ and $f(X) \rightarrow \{RU-gov, UA\}$ to identify whether a user is predominantly exposed to Russian pro-government (RU-gov), Russian independent (RU-ind) or Ukrainian (UA) news sources. 
Note that we remove all tweets that contain headlines of news and all retweets when constructing the language model of a user $X$. This ensures that we concentrate on the lingual choices in the tweets originating from the user, rather than the messages/sources that he/she (re)tweets. Also, as in the previous section, we only consider Russian-language tweets from Russian-speaking Twitter users, to ensure a fair comparison across all sides of the conflict. However, note that a large number of Ukrainian tweets are also in Russian, and our dataset contains representation from all three parties.

\vspace{-2em}
\begin{table}
\centering
\normalsize
\begin{tabular}{ |l | c | c | c |}
 \hline
  \textbf{Metric} & \specialcell{\textbf{RU-gov} vs. \\ \textbf{RU-ind}} & \specialcell{\textbf{RU-gov} vs. \\ \textbf{UA}} 
			& \specialcell{\textbf{RU-gov} vs. \\ \textbf{RU-ind} vs. \textbf{UA}} \\ \hline
  Precision & 0.66 & 0.71 & 0.52\\ \hline
  Recall & 0.66 & 0.70 & 0.52\\\hline
  Accuracy & 0.66 & 0.70 & 0.52 \\\hline
\end{tabular}
\vspace{1em}
\caption{\textbf{Predicting partisan slant in Twitter.} The results of a supervised machine learning experiment to identify partisan slant among Russian-speaking Twitter users during the Ukrainian conflict.}
\label{tab:twitter-prediction-performance}
\vspace{-3em}
\end{table}

\reftab{tab:twitter-prediction-performance} presents the averaged results of the 10-fold cross-validation of the proposed model. We note that the model has good performance, achieving an average accuracy of $70$\% and $66$\% in distinguishing between the users with RU-gov vs. UA slant and between the users with RU-gov vs. RU-ind slant, correspondingly. Comparing these results with the results for predicting the slant of news articles (Table~\ref{tab:articles-prediction-performance}), we note that the performance of \emph{RU-gov vs. RU-ind} classifier is comparable between the two cases whereas the performance of \emph{RU-gov vs. UA} classifier as well as the three-class classifier (accuracy of 52\% in comparison to expected 33\% for a random baseline) is slightly lower for the Twitter case. This can be probably attributed to the fact that the size of the text piece in an average news article is larger than in a collection of ten 140-character-long posts collected for a median Twitter user in our training and testing sets and, so, inferring political leaning from Twitter posts seems to be a harder problem than it is for news articles.  

\vspace{-1em}
\section{Conclusions and Discussion}
\vspace{-1em}
In this study we investigated the linguistic indicators of partisan confrontation in mainstream and social media during times of political upheaval by analyzing a dataset of more than 40M tweets and more than 4M news articles from the mass protests in Ukraine during 2013-2014 — known as "Euromaidan" — and the post-Euromaidan conflict between Russian, pro-Russian and Ukrainian forces in eastern Ukraine and Crimea. We designed a natural language processing algorithm to analyze at scale the linguistic markers which point to a particular political leaning in online media and showed that partisan slant in news articles can be automatically inferred with an accuracy of $60$--77\%. This difference in language in traditional news media is reflected in the word choices made by the supporters of similar partisan bent on Twitter — those who tweet news sources of a particular slant can be identified with an up to $70$\% accuracy based on their lingual choices in tweets other than the retweets of particular news sources. Our results have two implications: 

First, our results  contribute to the debate on the role of lingual choices in traditional and social media in fostering political frames and partisan discourse during political crises, and confirm that both traditional news sources and users on social media are identifiably partisan. It would be interesting to conduct a more general study into whether the reinforcing partisan nature of the discourse and the political divisions we observe arise from a general lack of empathy and trust in conflict situations, or whether this is specific to the Ukraine conflict.

Second, the results reveal the extent to which partisan rhetoric and political leanings can be automatically inferred from lingual choices, which has implications for the use of social media as a safe platform for free speech in dangerous conflicts. Furthermore, we are able to infer all this with only a ``coarse-grained'' approach: The party labels (i.e., pro-Ukrainian, Russian pro-government or Russian independent) are assigned to the news articles (or social media profiles) based on the polarity of the news agencies they originate from/retweet. The advantage of this approach is that it requires minimal manual efforts — it only requires to label the political slant of a number of news agencies (such as the 15 considered in this paper). However, the polarity and partisan rhetoric of articles may also vary between different topics and authors within a single news source, which we do not directly account for. We partially address this concern in the current paper by focusing only on the articles related to the Ukrainian events which are known for highly polarized rhetoric in both Russian and Ukrainian online media \cite{szostek2014media,zhukov2016reporting}. Although we settled on the coarse-grained approach as a proof-of-concept model, it is clear that a more fine-grained approach could allow for greater accuracy in identifying partisan tweets and political leanings of users and news media.

\vspace{-1em}
\section*{Acknowledgements}

The ‘A Shared Space and A Space for Sharing’ project (Grant No. ES/M00354X/1) is one of several funded through the EMoTICON network, which is funded through the following cross-council programmes: Partnership for Conflict, Crime and Security Research (led by the Economic and Social Research Council (ESRC)), Connected Communities (led by the Arts and Humanities Research Council (AHRC)), Digital Economy (led by the Engineering and Physical Sciences Research Council (EPSRC)). We would also like to thank the developers of the Kobzi application for providing the News-UA dataset. 

\bibliographystyle{splncs03}
\bibliography{biblio}

\end{document}